\documentclass[10pt,twocolumn,a4paper,conference]{IEEEtran}
\usepackage[english]{babel} % {english,greek}
\newcommand{\B}[1]{\boldsymbol{#1}}

\usepackage{amsmath}
\usepackage{amssymb}
\usepackage{epsfig}
\usepackage{psfrag}
\usepackage{algorithm}
\usepackage{algorithmic}

\DeclareMathOperator*{\argmax}{argmax}

\begin{document}
\selectlanguage{english}
\title{{ Multiple Parameter Estimation With Quantized Channel Output}}
\author{ Amine Mezghani $^1$, Felix Antreich $^2$  and Josef A. Nossek $^1$\\
$^1$Institute for Circuit Theory and Signal Processing, Technische Universit\"at M\"unchen, 80290 Munich, Germany\\
$^2$German Aerospace Center (DLR), Institute for Communications and Navigation, 82234 Wessling, Germany\\
Email:  $^1$\{Mezghani, Nossek\}@nws.ei.tum.de, $^2$felix.antreich@dlr.de}
%\name{\normalsize Author(s) Name(s)\thanks{Thanks to XYZ agency for funding.}}
%\address{\normalsize Author Affiliation(s)\\ {\normalsize \tt author@domain.org}}

\maketitle
\begin{abstract}
We present a general problem formulation for  optimal parameter estimation based on quantized observations, with application to antenna array communication and processing (channel estimation, time-of-arrival (TOA) and direction-of-arrival (DOA) estimation). The work is of interest in the case when low resolution A/D-converters (ADCs) have to be used to enable higher sampling rate and to simplify the hardware. An Expectation-Maximization (EM) based algorithm is proposed for solving this problem in a general setting. Besides, we derive the Cram\'er-Rao Bound (CRB) and discuss the effects of quantization and the optimal choice of the ADC characteristic. Numerical and analytical analysis reveals that reliable estimation may still be possible even when the quantization is very coarse.

%For illustration, a 2$\times$2 channel estimation using 1-bit ADC is considered in this extended abstract.     \\
\end{abstract}
\section{Introduction}
\label{section:introduction}
%%%%%%%%%%%%%%%%%%%%%%%%%%%%%%%%%%%%%%%%%%%%%%%%%%%%%%%%%%%%%%%%%%%%%%%%%%%%%%%%%%%%%%%%
In multiple-input
multiple-output (MIMO) communication systems, where  low power and low cost are key requirements, it is desirable to reduce the ADC resolution in order to save power and chip area \cite{schreier}. In fact, in high speed systems the sampling/conversion power may reach values in the order of the processing power. Therefore, coarse analog-to-digital converters (ADCs) may  be a cost-effective solution   for such applications, especially when the array size becomes very large or when the sampling rate becomes very high (in the GHz range) \cite{wentzloff}. Naturally, this generates a need for developing new detection and estimation algorithms operating on quantized data.
\par  An early work on the subject of estimating unknown parameters based on quantized can be found in \cite{curry}. In \cite{lok,ivrlac}, the authors studied channel estimation based on single bit quantizer (comparator). In this work, a more general setting for parameter estimation based on quantized observations will be studied, which covers many processing tasks, e.g. channel estimation, synchronization, delay estimation, Direction Of Arrival (DOA) estimation, etc. An Expectation Maximization (EM) based algorithm is proposed to solve the Maximum a Posteriori  Probability (MAP) estimation  problem. Besides, the Cram\'er-Rao Bound (CRB) has been derived to analyze the estimation performance and its behavior with respect to the signal-to-noise ratio (SNR). The presented results treat both cases: pilot aided and non-pilot aided estimation. We extensively deal with the extreme case of single bit quantized (comparator) which simplifies the sampling hardware considerably. We also focus on MIMO channel estimation and delay estimation as application area of the presented  approach.  Among others, a 2$\times$2 channel estimation using 1-bit ADC is considered, which shows that reliable estimation may still be possible even when the quantization is very coarse. In order to ease the theoretical derivations, we restrict
ourselves to real-valued systems. However,
the results can be easily extended and applied to 
complex valued-channels as we will do in Section~\ref{section:GNSS}.  
\par Our paper is organized as follows. Section \ref{section:scmodel} describes the general system model. In Section \ref{em_algo}, the EM-algorithm operating on quantized data is derive and  the estimation performance limit based on the Cram\'er-Rao Bound (CRB) is analyzed.   In Section \ref{section:ExampleI}, we deal with the single-input  single-output (SISO) channel estimation problem as a first application, then we generalize  the analysis to the multiple-antennas (MIMO) case in Section \ref{section:MIMO}. Finally we handle the problem of signal  quantization in the context of  Global Navigation Satellite Systems (GNSS) in Section \ref{section:GNSS}.
% The following work is a complementary work to \cite{mezghani2007,mezghanissd2007,mezghani_ICASSP2008}, where we have modified the conventional %linear and non-linear receiver designs, namely the  \emph{minimum mean square error} (MMSE) linear receiver  and the MMSE-Decision feedback %receiver, to take into account the presence of the quantizer in a non-empirical way.     
\par \textit{Notation:}  Vectors and matrices are denoted by lower and
upper case italic bold letters. The operators $(\bullet)^\mathrm {T}$, $(\bullet)^\mathrm {H}$, $\textrm{tr}(\bullet)$, 
$(\bullet)^*$, $\text{Re}(\bullet)$ and $\text{Im}(\bullet)$ stand for transpose, Hermitian transpose, trace of a matrix, complex conjugate, real and imaginary parts of a complex number, respectively.  $\textbf{\rmfamily{I}}_{M}$   denote the ($M \times M$) identity matrix. $\boldsymbol{x}_i$ is the $i$-th column of a  matrix $\B{X}$ and $x_{i,j}$ denotes the ($i$th, $j$th) element of it. 
The operator $\textrm{E}_{s|q}[\bullet]$ stands for expectation with respect to the random variable $s$ given $q$. The functions $p(s,q)$ and $p(s|q)$  symbolize the joint distribution and the conditional distribution of $s$ and $q$, respectively. Unless otherwise noted, all integrals are taken from $-\infty$ to $+\infty$. Finally, $\stackrel{\rm 1-bit}=$ symbolizes that the equality holds for the single bit case. 

\section{System Model}
\label{section:scmodel}
\begin{figure}[h]
\centerline{
\psfrag{ber}[c][c]{\small{Uncoded BER}}
\psfrag{pn}[c][c]{$\!\!\!\!\!\!\!\!\!\!\!\!\!\mathcal N(0,\sigma_\eta^2)$}
\psfrag{px}[c][c]{$p(\B{x})$}
\psfrag{N()}[c][c]{}
\psfrag{th}[c][c]{$\B{\theta}$}
\psfrag{x}[c][c]{$\B{x}$}
\psfrag{y}[c][c]{$\B{y}$}
\psfrag{z}[c][c]{$\B{r}$}
\psfrag{n}[c][c]{$\B{\eta}$}
\psfrag{pth}[c][c]{$p_\theta(\B{\theta})$}
\psfrag{Q}[c][c]{$ \mathbb{I}_{D(\B{r})}(\B{y})$}
\psfrag{f(x,th)}[c][c]{$f(\B{x},\B{\theta})$}
\epsfig{file=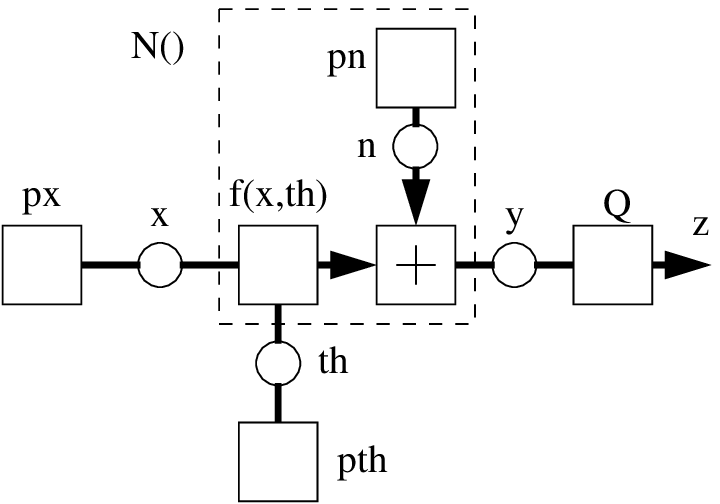, width =7.5cm}}
\caption{Factor graph representation.}
\label{channel_graph}
\end{figure}

%\cite{ref1}\cite{ref2}\cite{ref3}\cite{wiesel04multiuser}\cite{schubert}\cite{palomar}
%%%%%%%%%%%%%%%%%%%%%%%%%%%%%%%%%%%%%%%%%%%%%%%%%%%%%%%%%%%%%%%%%%%%%%%%%%%%%%%%%%%%%%%%
As mentioned before, we start from a general signal model, described by:
\begin{equation}
\B{r}=Q(\B{y}), ~~{\rm with}
\end{equation}
\begin{equation}
\B{y}= \B{f}(\B{x},\B{\theta})+\B{\eta},
\end{equation}
where $\B{y}$ is the unquantized receive vector of dimension $N$, $\B{f}(\cdot,\cdot)$ is a general multidimensional system function of the unknown parameter vector $\B{\theta}$, to be estimated, and the known or unknown data vector $\B{x}$, while $\B{\eta}$ is an i.i.d. Gaussian noise with variance $\sigma_\eta^2$ in each dimension. We assume that the noise variance $\sigma_\eta^2$ is known, although this part of the  work can be easily  extended to the case where $\sigma_\eta^2$ is part of $\B{\theta}$. The operator $Q(\cdot)$ represents the quantization process, where each component $y_i$ is mapped to a quantized value from a finite set of code words as follows
\begin{equation}
r_i=Q(y_i), ~~{\rm if}~~y_i\in [r_i^{\rm lo}, r_i^{\rm up}).
\end{equation}
 Thereby $r_i^{\rm lo}$  and $r_i^{\rm up}$ are the lower value and the upper limits associated to quantized value $r_i$. Additionally, we denote the prior distribution of the parameter vector by $p_{\theta} (\B{\theta})$ when available. Similarly the prior $p_x(\B{x})$  is also known, and can  for instance be obtained from the extrinsic information of the decoder output. The joint probability density function involving all random system variables reads consequently  as 
\begin{equation}
\begin{aligned}
p(\B{r},\B{y},\B{x},\B{\theta})&=  \mathbb{I}_{D(\B{r})}(\B{y})\frac{1}{(2\pi)^{\frac{N}{2}} \sigma_\eta^{N}} {\rm e}^{-\frac{\left\| \B{y}-\B{f}(\B{\theta},\B{x})\right\|^2_2}{2\sigma_\eta^2}}p_x(\B{x})p_\theta(\B{\theta}),
\end{aligned}
\label{joint_pdf}
\end{equation}
where $\mathbb{I}$ denotes the indicator function taking one if 
\begin{equation}
\begin{aligned}
\B{y} \in  D(\boldsymbol{r})=\left\{\boldsymbol{y}\in \mathbb {R}^N |r_i^{\rm lo}\leq y_{i} \leq r_i^{\rm up}; \forall i\in \{1,\ldots, N\}\right\},
 \end{aligned}
\end{equation}
and 0 otherwise. Note that this special factorization of the joint density function is crucial for solving and analyzing the estimation problem. A factor graph representation of the joint probability density is  given in Fig.~\ref{channel_graph} to illustrate this property. Each random variable is represented by a circle, referred to as variable node, and each factor of the global function (\ref{joint_pdf})  corresponds to a square, called functional node or factor node.
%%%%%%%%%%%%%%%%%%%%%%%%%%%%%%%%%%%%%%%%%%%%%%%%%%%%%5
\section{Construction of the Estimation Algorithm and Performance Bound}
\label{em_algo}
%%%%%%%%%%%%%%%%%%%%%%%%%%%%%%%%%%%%%%%%%%%%%%%%%%%%
Given the quantized observation $\B{r}$, and the log-likelihood function
\begin{equation}
\begin{aligned}
\mathcal{L}(\B{\theta})=\ln \int \int p(\B{r},\B{y},\B{x},\B{\theta}) {\rm d}\B{x} {\rm d}\B{y}=\ln p(\B{r},\B{\theta}),
\end{aligned}
\end{equation}
 our goal is to find the MAP estimate $\hat{\B{\theta}}$ given by
\begin{equation}
\begin{aligned}
 \hat{\B{\theta}}=\argmax_{\B{\theta}}  \mathcal{L}(\B{\theta}).
\end{aligned}
\label{MAP_MAX}
\end{equation}
Naturally, the MAP solution $\hat{\B{\theta}}$ has to satisfy \\
\begin{equation}
\begin{aligned}
 \nabla_{\B{\theta}}\mathcal{L}(\B{\theta})=0.
\end{aligned}
\end{equation}
This condition can be written as:
\begin{align}
 \nabla_{\B{\theta}}\mathcal{L}(\B{\theta})&= \frac{\nabla_{\B{\theta}} p(\B{r},\B{\theta})   }{ p(\B{r},\B{\theta})} \nonumber \\
 &=\int \int \frac{\nabla_{\B{\theta}} p(\B{r},\B{y},\B{x},\B{\theta})   }{ p(\B{r},\B{\theta})} {\rm d}\B{x} {\rm d}\B{y} \nonumber \\
 &=\int \int \frac{\nabla_{\B{\theta}} p(\B{r},\B{y},\B{x},\B{\theta})   }{ p(\B{r},\B{\theta})} \cdot \frac{p(\B{x},\B{y}|\B{r},\B{\theta})}{p(\B{x},\B{y}|\B{r},\B{\theta})} {\rm d}\B{x} {\rm d}\B{y} \nonumber \\
 &=\int \int \frac{\nabla_{\B{\theta}} p(\B{r},\B{y},\B{x},\B{\theta})   }{ p(\B{r},\B{y},\B{x},\B{\theta})} \cdot p(\B{x},\B{y}|\B{r},\B{\theta}) {\rm d}\B{x} {\rm d}\B{y} \nonumber \\
 &= {\rm E}_{\B{x},\B{y}|\B{r},\B{\theta}} \left[ \nabla_{\B{\theta}} \ln p(\B{r},\B{y},\B{x},\B{\theta}) \right] \stackrel{!}=0.
\label{KKT_cond}
\end{align}
There is also another way to write the optimality condition, by first integrating out the variable $\B{y}$ to obtain the conditional probability of the quantized received vector
\begin{equation}
\begin{aligned}
p(\B{r}|\B{x},\B{\theta})&= \int_{r_i^{\rm lo}}^{r_i^{\rm up}} \frac{1}{(2\pi)^{\frac{N}{2}} \sigma_\eta^{N}} {\rm e}^{-\frac{\left\| \B{y}-\B{f}(\B{\theta},\B{x})\right\|^2_2}{2\sigma_\eta^2}}{\rm d}\B{y} \\
&=\prod_i ( \Phi(\frac{r_i^{\rm up}-f_i(\B{x}, \hat{\B{\theta}})}{\sigma_\eta} )-\Phi(\frac{r_i^{\rm lo}-f_i(\B{x}, \hat{\B{\theta}})}{\sigma_\eta} )),
\end{aligned}
\end{equation}
where $\Phi(x)$ represents the cumulative Gaussian distribution reading as
\begin{equation}
\Phi(x)=\frac{1}{\sqrt{2\pi}}\int_{-\infty}^x\exp(-t^2/2) \rm{d}t.
\end{equation}
Therefore, we also obtain an alternative condition as
\begin{align}
 \nabla_{\B{\theta}}\mathcal{L}(\B{\theta})
 &=\int \frac{\nabla_{\B{\theta}} \int  p(\B{r},\B{y},\B{x},\B{\theta}){\rm d}\B{y}   }{ p(\B{r},\B{\theta})} {\rm d}\B{x} \nonumber  \\
 &= \int \frac{\nabla_{\B{\theta}} p(\B{r},\B{x},\B{\theta})   }{ p(\B{r},\B{\theta})} \cdot \frac{p(\B{x}|\B{r},\B{\theta})}{p(\B{x}|\B{r},\B{\theta})} {\rm d}\B{x} \nonumber \\
 &=\int \frac{\nabla_{\B{\theta}} p(\B{r},\B{x},\B{\theta})   }{ p(\B{r},\B{x},\B{\theta})} \cdot p(\B{x}|\B{r},\B{\theta}) {\rm d}\B{x}  \nonumber \\
 &= {\rm E}_{\B{x}|\B{r},\B{\theta}} \left[ \nabla_{\B{\theta}} \ln p(\B{r},\B{x},\B{\theta}) \right] \stackrel{!}=0,
 \label{KKT_cond2}
\end{align}
which can be explicitly written as,
 \begin{equation}
\begin{aligned}
&\!\!\!\!\!\sum_i\! {\rm E}_{\B{x}|\B{r},\hat{\B{\theta}}}\!\!\!\left[\!\frac{({\rm e}^{-\frac{(r_i^{\rm up}-f_i(\B{x}, \hat{\B{\theta}}))^2}{2\sigma_\eta^2} }\!\!-\!{\rm e}^{-\frac{(r_i^{\rm lo}-f_i(\B{x}, \hat{\B{\theta}}))^2}{2\sigma_\eta^2} })\nabla_{\!\B{\theta}}\! f_i(\B{x}, \hat{\B{\theta}}) }{ \sqrt{2\pi} \sigma_\eta ( \Phi(\frac{r_i^{\rm up}-f_i(\B{x}, \hat{\B{\theta}})}{\sigma_\eta} )-\Phi(\frac{r_i^{\rm lo}-f_i(\B{x}, \hat{\B{\theta}})}{\sigma_\eta} ))}\!\right]\\
&\hspace{6.0cm}-\frac{\nabla_{\B{\theta}}p_\theta(\hat{\B{\theta}})}{p_\theta(\hat{\B{\theta}})}=0 \\
&\!\!\!\!\!\stackrel{\rm 1-bit}=-\sum_i\! {\rm E}_{\B{x}|\B{r},\hat{\B{\theta}}}\!\!\!\left[r_i \frac{{\rm e}^{-\frac{f_i(\B{x}, \hat{\B{\theta}})^2 }{2 \sigma_\eta^2}}\nabla_{\!\B{\theta}}\! f_i(\B{x}, \hat{\B{\theta}})}{\sqrt{2\pi}\sigma_\eta\Phi(\frac{r_if_i(\B{x}, \hat{\B{\theta}})  }{ \sigma_\eta})}\right]\!-\frac{\nabla_{\B{\theta}}p_\theta(\hat{\B{\theta}})}{p_\theta(\hat{\B{\theta}})}=0,
\end{aligned}
\label{KKT}
\end{equation}
where the last step holds for the single bit case, i.e. $r_i \in \{\pm 1\}$.
%%%%%%%%%%%%%%%%%%%%%%%%%%%%%%%%%%%%%%%%%%%%%%%%%%%%%5
\subsection{EM-Based MAP Solution}
\label{subsec:EM}
%%%%%%%%%%%%%%%%%%%%%%%%%%%%%%%%%%%%%%%%%%%%%%%%%%%%
In general, solving (\ref{KKT}) is intractable, thus we resort to the popular Expectation Maximization (EM) algorithm as iterative procedure for   solving the condition (\ref{KKT_cond}) in the following recursive way
\begin{equation}
\begin{aligned}
 {\rm E}_{\B{x},\B{y}|\B{r},\B{\theta}^l} \left[ \nabla_{\B{\theta}} \ln p(\B{r},\B{y},\B{x},\B{\theta}^{l+1}) \right] =0.
\end{aligned}
\end{equation}
Thus, at each iteration $l$ the following two steps are performed: \vspace{0.15cm} \\
\underline{E-step:} Compute the expectation\\
\begin{equation}
\begin{aligned}
   &g(\B{r}, \B{\theta} ,\hat{\B{\theta}}^{l})= 
{\rm E}_{\B{x},\B{y}|\B{r},\hat{\B{\theta}}^{l}} [\ln p(\B{r},\B{y},\B{x},\B{\theta})] +  {\rm const} \\
&~~={\rm E}_{\B{x},\B{y}|\B{r},\hat{\B{\theta}}^{l}} \Big[2\B{y}^{\rm T}\B{f}(\B{\theta},\B{x})\!+\!\left\| \B{f}(\B{\theta},\B{x})\right\|^2_2  \Big]/(2\sigma_\eta^2) \!+ \!\ln p_\theta(\B{\theta}) \\
&~~={\rm E}_{\B{x}|\B{r},\hat{\B{\theta}}^{l}} \Big[2( f_i(\hat{\B{\theta}}^l,\B{x})  +{\rm E} [\B{\eta}|\B{x},\B{r},\hat{\B{\theta}}^{l}] ) ^{\rm T}\B{f}(\B{\theta},\B{x})- \\
&~~~~~~\left\| \B{f}(\B{\theta},\B{x})\right\|^2_2  \Big]/(2\sigma_\eta^2)  + \ln p_\theta(\B{\theta}), 
\end{aligned}
\label{estep}
\end{equation}
where            
\begin{equation}
\begin{aligned}
{ \rm E}[\eta_i|\B{x},\B{r},\hat{\B{\theta}}^{l}]&= -\frac{\sigma_\eta}{\sqrt{2\pi}}\cdot \frac{{\rm e}^{-\frac{(r_i^{\rm up}-f_i(\B{x}, \hat{\B{\theta}}^{l}))^2  }{ 2\sigma_\eta^2}}-{\rm e}^{-\frac{(r_i^{\rm lo}-f_i(\B{x}, \hat{\B{\theta}}^{l}) )^2 }{ 2\sigma_\eta^2 }}}{\Phi(\frac{r_i^{\rm up}-f_i(\B{x}, \hat{\B{\theta}}^{l})  }{ \sigma_\eta})-\Phi(\frac{r_i^{\rm lo}-f_i(\B{x}, \hat{\B{\theta}}^{l})  }{ \sigma_\eta})} \\
&\stackrel{\rm 1-bit}= r_i\frac{\sigma_\eta}{\sqrt{2\pi}}\cdot \frac{{\rm e}^{-\frac{f_i(\B{x}, \hat{\B{\theta}}^{l})^2  }{2 \sigma_\eta^2}}}{\Phi(\frac{r_if_i(\B{x}, \hat{\B{\theta}}^{l})  }{ \sigma_\eta})}.
\nonumber 
\end{aligned}
\end{equation}
\underline{M-step:} Solve the maximization
\begin{equation}
\begin{aligned}
 \hat{\B{\theta}}^{l+1}=\argmax_{\B{\theta}}  g(\B{r}, \B{\theta} ,\hat{\B{\theta}}^{l}).
\end{aligned}
\label{mstep}
\vspace{0.25cm}
\end{equation}
In many cases, this maximization is much easier than (\ref{MAP_MAX}), as we will see in the examples considered later.
%%%%%%%%%%%%%%%%%%%%%%%%%%%%%%%%%%%%%%%%%%%%%%%%%%%%%%%%%%%%%%%%%%%%%%%%%%%%%%%%%%%%%%%%
\subsection{Standard Cram\'er-Rao Bound (CRB)}
\label{SEC_CRB}
The standard CRB\footnote{ The standard CRB, in contrast to the Bayesian CRB, holds for a deterministic parameter, i.e. the prior $p_\theta(\B{\theta})$ is not taken into account.} is the lower bound on the estimation error for any unbiased estimator, that can be obtained from the Fisher information matrix $\B{J}(\B{\theta})$ under certain conditions
\begin{equation}
\begin{aligned}
{\rm E}[(\B{\theta}-\hat{\B{\theta}})(\B{\theta}-\hat{\B{\theta}})^{\rm T}] \succeq (\B{J}(\B{\theta}))^{-1}.
\end{aligned}
\end{equation}
Hereby, the Fisher information matrix reads as \cite{papoulis}
\begin{equation}
\begin{aligned}
\! \B{J}\!&= {\rm E}_{\B{r}|\B{\theta}} [\nabla_{\B{\theta}}\mathcal{L}(\B{\theta})\nabla_{\B{\theta}}^{\rm T}\mathcal{L}(\B{\theta})]    \\
&= {\rm E}_{\B{r}|\B{\theta}} \Big[  {\rm E}_{\B{x}|\B{r},\B{\theta}} \left[ \nabla_{\B{\theta}} \ln p(\B{r},\B{x},\B{\theta}) \right] \cdot \\
&~~~~~~~~~~{\rm E}_{\B{x}|\B{r},\B{\theta}} \left[ \nabla_{\B{\theta}}^{\rm T} \ln p(\B{r},\B{x},\B{\theta}) \right] \Big]\\
%&= \sum_i{\rm E}_{\B{r}|\B{\theta}} \Bigg[  {\rm E}_{\B{x},y_i|\B{r},\B{\theta}} \left[ \frac{y_i-f_i(\B{x},\B{\theta})}{\sigma_\eta^2} \nabla_{\B{\theta}} f_i(\B{x},\B{\theta}) \right] \cdot \\
%&~~~~~~~~~~~~~~~~{\rm E}_{\B{x},y_i|\B{r},\B{\theta}}\! \left[\frac{y_i-f_i(\B{x},\B{\theta})}{\sigma_\eta^2} \nabla_{\B{\theta}}^{\rm T} f_i(\B{x},\B{\theta})  \right]  \Bigg]\\
&= \!\!\!\sum_i\!{\rm E}_{\B{r}|\B{\theta}} \!\left[ \! {\rm E}_{\B{x}|\B{r},\B{\theta}} \!\Bigg[\!\frac{{\rm e}^{-\frac{(r_i^{\rm up}\!-\!f_i(\B{x}, \B{\theta}))^2}{2\sigma_\eta^2} }\!\!\!-\!{\rm e}^{-\frac{(r_i^{\rm lo}\!-\!f_i(\B{x}, \B{\theta}))^2}{2\sigma_\eta^2} } }{   \Phi(\frac{r_i^{\rm up}\!-\!f_i(\B{x}, \B{\theta})}{\sigma_\eta} )\!-\!\Phi(\frac{r_i^{\rm lo}\!-\!f_i(\B{x}, \B{\theta})}{\sigma_\eta} )}\! \nabla_{\B{\theta}}^{\rm T} \!f_i(\B{x},\B{\theta}) \! \Bigg] \! \cdot  \right. \\
&~~\left.{\rm E}_{\B{x}|\B{r},\B{\theta}} \!\Bigg[\!\frac{{\rm e}^{-\frac{(r_i^{\rm up}\!-\!f_i(\B{x}, \B{\theta}))^2}{2\sigma_\eta^2} }\!\!\!-\!{\rm e}^{-\frac{(r_i^{\rm lo}\!-\!f_i(\B{x}, \B{\theta}))^2}{2\sigma_\eta^2} } }{   \Phi(\frac{r_i^{\rm up}\!-\!f_i(\B{x}, \B{\theta})}{\sigma_\eta} )\!-\!\Phi(\frac{r_i^{\rm lo}\!-\!f_i(\B{x}, \B{\theta})}{\sigma_\eta} )} \nabla_{\B{\theta}}^{\rm T} f_i(\B{x},\B{\theta})\!    \Bigg] \right]\frac{1}{2\pi\sigma_\eta^2}.
%%%%%%%%%%%%%%%%%%%%%%
%&\! \!\sum_{\B{r},i}\! {\rm E}_{\B{x}|\B{r},\B{\theta}}\!\!\!\left[\!\frac{({\rm e}^{-\frac{(r_i^{\rm up}-f_i(\B{x}, \B{\theta}))^2}{2\sigma_\eta^2} }\!\!\!-\!{\rm e}^{-\frac{(r_i^{\rm lo}-f_i(\B{x}, \B{\theta}))^2}{2\sigma_\eta^2} })^2\nabla_{\!\B{\theta}}\! \nabla_{\!\B{\theta}}^{\rm T} f_i(\B{x}, \B{\theta}) }{ 2\pi \sigma_\eta^2 ( \Phi(\frac{r_i^{\rm up}-f_i(\B{x}, \B{\theta})}{\sigma_\eta} )-\Phi(\frac{r_i^{\rm lo}-f_i(\B{x}, \B{\theta})}{\sigma_\eta} ))^2}\!\right]\!.
\end{aligned}
\end{equation}
In the pilot-based estimation case ($\B{x}$ is known), it simplifies to
\begin{equation}
\begin{aligned}
\!\!\!\!\!\!\!\!\!\!\!\!\!\!\B{J}\!
&=  \sum_{i,r_i}   \!\frac{\!({\rm e}^{-\frac{(r_i^{\rm up}\!-\!f_i(\B{x}, \B{\theta}))^2}{2\sigma_\eta^2} }\!\!-\!{\rm e}^{-\frac{(r_i^{\rm lo}\!\!-\!f_i(\B{x}, \B{\theta}))^2}{2\sigma_\eta^2} })^2\nabla_{\!\B{\theta}}\! f_i(\B{x}, \B{\theta}) \nabla_{\!\B{\theta}}^{\rm T}\! f_i(\B{x}, \B{\theta})\!}{2 \pi \sigma_\eta^2 ( \Phi(\frac{r_i^{\rm up}-f_i(\B{x}, \B{\theta})}{\sigma_\eta} )-\Phi(\frac{r_i^{\rm lo}-f_i(\B{x}, \B{\theta})}{\sigma_\eta} ))}\\
& \stackrel{\rm 1-bit}= \sum_i   \frac{\!{\rm e}^{-\frac{(f_i(\B{x}, \B{\theta}))^2}{\sigma_\eta^2} }\!\!\nabla_{\!\B{\theta}}\! f_i(\B{x}, \B{\theta}) \nabla_{\!\B{\theta}}^{\rm T}\! f_i(\B{x}, \B{\theta})\!}{ 2\pi \sigma_\eta^2  \Phi(\frac{f_i(\B{x}, \B{\theta})}{\sigma_\eta} )\Phi(-\frac{f_i(\B{x}. \B{\theta})}{\sigma_\eta} )}.
\end{aligned}
\label{CRLB}
\end{equation}
Additionally, in the low SNR regime ($\sigma_\eta \gg |f_i(\B{x}, \B{\theta})|$), (\ref{CRLB}) can be approximated by
 \begin{equation}
\begin{aligned}
\B{J}\approx  \frac{\rho_Q}{ { \sigma_\eta^2}}  \sum_{i}  \nabla_{\!\B{\theta}}\! f_i(\B{x}, \B{\theta}) \nabla_{\!\B{\theta}}^{\rm T}\! f_i(\B{x}, \B{\theta}),
\end{aligned}
\label{fisher_low}
\end{equation}
where the factor
%\vspace{-0.8cm}
\begin{equation}
\begin{aligned}
\rho_Q = \frac{1}{2 \pi} \sum_{r}   \!\frac{\!({\rm e}^{-\frac{(r^{\rm up})^2}{2\sigma_\eta^2} }\!\!-\!{\rm e}^{-\frac{(r^{\rm lo})^2}{2\sigma_\eta^2} })^2   \!}{  \Phi(\frac{r^{\rm up}}{\sigma_\eta} )-\Phi(\frac{r^{\rm lo}}{\sigma_\eta} )} 
%\stackrel{\rm 1-bit}= \frac{2}{\pi} 
\leq 1,
\end{aligned}
\label{rho_Q}
\end{equation}
depends only on the quantizer characteristic (here assumed to be the same for all dimensions) and represents the information loss compared to the unquantized case at low SNR, in the pilot-based estimation case. For the single bit case, i.e., $(r^{\rm lo},r^{\rm up})\in\{(-\infty,0),(0,\infty)\}$, the Fisher information loss $\rho_Q$ is equal to $2/\pi$, which coincides to the result found in \cite{mezghaniisit2007,mezghaniisit2009}  in terms of the Shannon's mutual information of the channel. For the case that we use a uniform symmetric mid-riser type quantizer \cite{proaksis}, the quantized receive alphabet is given by
\begin{equation}
  r_{i}\in \{ (-\frac{2^b}{2}-\frac{1}{2}+k)\Delta;\textrm{ } k=1,\cdots,2^b\}=\mathcal{R},  
\end{equation}    
where $\Delta$ is the quantizer step-size and $b$ is the number of quantizer bits. Hereby the lower and upper quantization thresholds  are
\begin{equation*}
r_{i}^{\rm lo }=
\begin{cases}
r_{i}-\frac{\Delta}{2} & \mathrm{for} \quad r_{i}\geq -\frac{\Delta}{2}(2^b-2)\\
-\infty & \mathrm{otherwise,} 
\end{cases}
\end{equation*}
and 
\begin{equation*}
r_{i}^{\rm up }=
\begin{cases}
r_{i}+\frac{\Delta}{2} & \mathrm{for} \quad r_{i}\leq \frac{\Delta}{2}(2^b-2)\\
+\infty & \mathrm{otherwise}.
\end{cases}
\end{equation*}
In order to optimize the Fisher information  at low SNR (\ref{fisher_low}) and get close to the full precision estimation performance, we need to maximize $\rho_Q$ from (\ref{rho_Q}) with respect to the quantizer characteristic. Table~\ref{uniform} shows the optimal (non-unform) step size $\Delta_{\rm opt}$ (normalized by $\sigma_\eta^2$) of the uniform quantizer described above, which maximizes $\rho_Q$ for $b\in\{1,2,3,4\}$. If we do not restrict the characteristic to be uniform, then we get the optimal quantization thresholds which  maximize $\rho_Q$  in Table~\ref{nonuniform}. We note that the obtained uniform/non-uniform quantizer optimized in terms of the estimation performance is not equivalent to the optimal quantizer, which we would get when minimizing the distortion, for a Gaussian input \cite{proaksis}. In addition, contrary to the quantization for minimum distortion the performance gap between the uniform and non-uniform quantization in our case is quite insignificant, as we can see from both tables.   
\begin {table}[thp]%
\caption {Optimal Uniform Quantizer.}
\label{uniform}\centering %
\begin{tabular}{ccc}
\hline %
$b$ & $\Delta_{\rm opt}$ & $\rho_Q$ \\
\hline %
1 & - & $2/\pi$  \\\hline %
2 &0.704877 & 0.825763  \\\hline %
3 &0.484899 & 0.945807  \\\hline %
4 &0.294778 & 0.984735  \\\hline %
\end {tabular}
\end {table}
\begin {table}[thp]%
\caption {Optimal non-Uniform Quantizer.}
\label {nonuniform}\centering %
\begin{tabular}{ccc}
\hline %
$b$ & Optimal thresholds  & $\rho_Q$ \\
\hline %
1 & 0 & $2/\pi$  \\\hline %
2 &0;$\pm$0.704877 & 0.825763  \\\hline %
3 &0;$\pm$0.306654;$\pm$0.895107;$\pm$1.626803 & 0.956911  \\\hline %
4 &0;$\pm$0.143878;$\pm$0.4204768;$\pm$0.708440;$\pm$1.017896; & 0.989318 \\
 &             $\pm$1.364802;$\pm$1.780058;$\pm$2.346884  &
\\\hline %
\end {tabular}
\end {table}    
\par In the following, the theoretical findings will be applied to the channel estimation problem and for a GNSS problem with quantized observations.
%%%%%%%%%%%%%%%%%%%%%%%%%%%%%%%%%%%%%%%%%%%%%%%%%%%%%%%%%%%%%%%%%%%%%%%%%%%%%%%%%%%%%%%%%%%%%%
%%%%%%%%%%%%%%%%%%%%%%%%%%%%%%%%%%%%%%%%%%%%%%%%%%%%%%%%%%%%%%%%%%%%%%%%%%%%%%%%%%%%%%%%%%%%%%%
\section{Example 1:  SISO channel estimation}
\label{section:ExampleI}
We first review the  simple problem of SISO channel estimation considered in \cite{ivrlac}.
\subsection{Pilot-based single bit estimation (one-tap channel)}  
\begin{figure}[ht]
\centerline{
\psfrag{Fisher Information}[c][c]{\footnotesize  Fisher Information$\cdot h^2$}
\psfrag{SNR}[c][c]{  \footnotesize $h^2/\sigma_\eta^2$ (linear)}
\epsfig{file=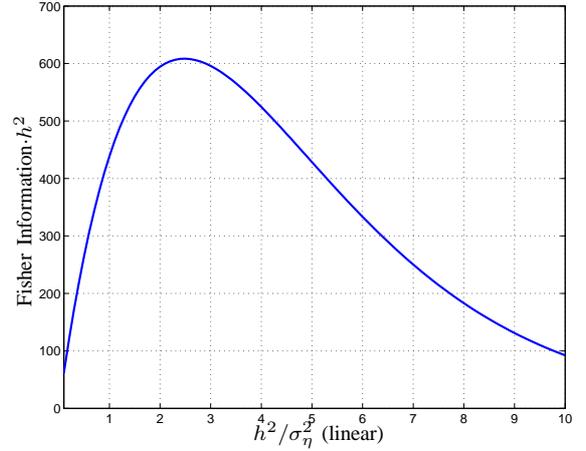, width =8.5cm}}
\caption{Fisher Information vs. $\sigma_\eta^2$ for a SISO channel, $b=1$, $N=200$.}
\label{fish}
\end{figure}

\begin{figure}[ht]
\centerline{
\psfrag{CRLB, MSE}[c][c]{\footnotesize  (CRB, MSE)/$h^2$}
\psfrag{SNR}[c][c]{ \footnotesize $h^2/\sigma_\eta^2$ (linear)} 
\epsfig{file=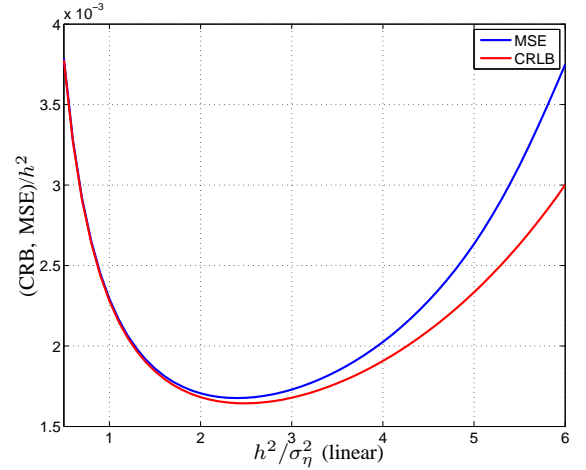, width =8.5cm}}
\caption{Estimation error vs. $\sigma_\eta^2$ for a SISO channel, $b=1$, $N=200$.}
\label{siso}
\end{figure}

The SISO one-tap channel model is given by
\begin{equation}
r_i=\textrm{sign}(y_i)=\textrm{sign}(h  x_i + \eta_i),\textrm{ for } i\in\{1,\ldots,N\},
\end{equation}
where $N$ is the pilot length and $x_i\in\{-1,1\}$ is the transmitted pilot sequence with normalized power.  The channel coefficient $h\in\mathbb{R}$ is  here our unknow parameter, i.e. $\B{\theta}=[h] $. \\
It can be shown by solving the optimality condition (\ref{KKT}) (with uniform prior $p_{\B{\theta}}(\B{\theta})$) that the ML-estimate of the scalar channel from the single bit outputs $r_i$, is given by \cite{ivrlac}
\begin{equation}
\hat{h}=\sqrt{2\sigma_\eta^2}\textrm{erf}^{-1}(\frac{\B{r}^{\rm T}\B{x}}{N}).
\end{equation}
Besides, the Fisher information (\ref{CRLB}) becomes in this case
\begin{align}
J(h)= \frac{N{\rm e}^{-\frac{h^2}{\sigma_\eta^2} }}{ 2\pi \sigma_\eta^2  \Phi(\frac{h}{\sigma_\eta} )\Phi(-\frac{h}{\sigma_\eta} )}.
\end{align}
This expression of the  Fisher information  is shown in Fig.~\ref{fish} for $N=200$ as function of $h^2/\sigma_\eta^2$.  In Fig.~\ref{siso} the CRB, i.e.  $1/J$ and the relative exact \emph{mean square error} (MSE) of the ML-estimate from $N=200$ observations, both normalized by $h^2$, are depicted as function of the SNR$=h^2/\sigma_\eta^2$.  We interestingly observe that above a certain SNR, the estimation performance degrades, which means that noise may be favorable at a certain level, contrary to the unquantized channel. This phenomenon is known as  stochastic resonance, which occurs when dealing  with such nonlinearities. We  can naturally seek the optimal SNR that maximizes the normalized  Fisher information, i.e. minimizes CRB$/h^2$:
  \begin{align}
\left.\frac{h^2}{\sigma_\eta^2}\right|_{\rm opt} \!\!= \!\argmax_ {\gamma} \! \frac{N{\rm e}^{-\gamma }}{ 2\pi  \gamma \Phi(\sqrt{\gamma})\Phi(-\sqrt{\gamma} )}= 2.4807 \equiv  3.9458{\rm dB}.
\end{align} 
This results obtained by numerical optimization of the  Fisher information coincide with the results found in \cite{ivrlac} through  observations at asymptotically large $N$. 
\vspace{-0.3cm}
\vspace{-0.5cm}
\subsection{Pilot Based Estimation (two-tap channel)}
\label{section:twotap}
Now let us consider a more general setting with a two-tap inter-symbol-interference (ISI) channel
\begin{equation}
r_i=\textrm{sign}(y_i)=\textrm{sign}(h_0  x_i + h_1 x_{i-1}+ \eta_i),\textrm{ for } i\in\{1,\ldots,N\},
\end{equation}
where $h_0$ and $h_1$ are the channel taps. Again we utilize a binary amplitude pilot sequence
$\B{x} \in \{-1, 1\}^N$ and we try to find the the ML-estimate of the parameter vector $\B{\theta}=[h_0, h_1]^{\rm T}$ in closed form.  Ignoring the first output $r_1$, the ML-condition (\ref{KKT}) turns to be ($p_\theta(\B{\theta})=1$)
\begin{equation}
\begin{aligned}
\sum_{i=2}^N r_ix_i\frac{{\rm e}^{-\frac{(h_0x_i+h_1x_{i-1})^2  }{2 \sigma_\eta^2}}}{\Phi(\frac{r_i(h_0x_i +h_1x_{i-1} )  }{ \sigma_\eta})}&=0, \\
\sum_{i=2}^N r_ix_{i-1}\frac{{\rm e}^{-\frac{(h_0x_i+h_1x_{i-1})^2  }{ 2\sigma_\eta^2}}}{\Phi(\frac{r_i(h_0x_i +h_1x_{i-1} )  }{ \sigma_\eta})}&=0. \\
\end{aligned}
\end{equation}
Taking the sum and the difference of these equations delivers respectively
\begin{equation}
\begin{aligned}
\sum_{i=2}^N r_i(x_i+x_{i-1})\frac{{\rm e}^{-\frac{(h_0x_i+h_1x_{i-1})^2  }{2 \sigma_\eta^2}}}{\Phi(\frac{r_i(h_0x_i +h_1x_{i-1} )  }{ \sigma_\eta})}&=0, \\
\sum_{i=2}^N r_i(x_i-x_{i-1})\frac{{\rm e}^{-\frac{(h_0x_i+h_1x_{i-1})^2  }{ 2\sigma_\eta^2}}}{\Phi(\frac{r_i(h_0x_i +h_1x_{i-1} )  }{ \sigma_\eta})}&=0. \\
\end{aligned}
\end{equation}
Next, we multiply the numerator and denominator of each equation by $\Phi(-\frac{r_i(h_0x_i +h_1x_{i-1} )  }{ \sigma_\eta})$ to get 
\begin{equation}
\begin{aligned}
\sum_{i=2}^N r_i(x_i+x_{i-1})\frac{\Phi(-r_ix_i\frac{h_0 +h_1   }{ \sigma_\eta})}{\Phi(\frac{h_0 +h_1   }{ \sigma_\eta})\Phi(-\frac{h_0 +h_1   }{ \sigma_\eta})}&=0, \\
\sum_{i=2}^N r_i(x_i-x_{i-1})\frac{\Phi(-r_ix_i\frac{h_0 +h_1   }{ \sigma_\eta})}{\Phi(\frac{h_0 -h_1  }{ \sigma_\eta})\Phi(-\frac{h_0 -h_1   }{ \sigma_\eta})}&=0. \\
\end{aligned}
\end{equation}
Then, using the fact that
\begin{equation}
\begin{aligned}
 2\Phi(-r_ix_i\frac{h_0 +h_1   }{ \sigma_\eta})=1- r_ix_i {\rm erf}(\frac{h_0 +h_1   }{\sqrt{2} \sigma_\eta}),
 \end{aligned}
\end{equation}
where ${\rm erf}(\cdot)$ denotes the Gaussian error function, we get
\begin{equation}
\begin{aligned}
\sum_{i=2}^N (x_i+x_{i-1})r_i&=(N-1+\sum_{i=2}^N x_ix_{i-1}) {\rm erf} (\frac{h_0+h_1}{\sqrt{2}\sigma_\eta}), \\
\sum_{i=2}^N  (x_i-x_{i-1})r_i&=(N-1-\sum_{i=2}^N x_ix_{i-1}) {\rm erf} (\frac{h_0-h_1}{\sqrt{2}\sigma_\eta}).
\end{aligned}
\end{equation}
Finally, solving the last equations with respect to $h_0$ and $h_1$, we get the ML solution.
\begin{equation}
\begin{aligned}
\!\!\!\hat{h}_{0} \!=  \!\sqrt{\!\frac{\sigma_\eta^2}{2}}\! \!\left(\! \! {\rm erf}^{-1}  \!\!\left[ \! \frac{\sum\limits_{i=2}^N \! (x_i\!+\!x_{i-1})r_i}{N\!+\!\sum\limits_{i=2}^N x_ix_{i-1} } \!\right] \! \!+  \! {\rm erf}^{-1}  \! \!\left[ \!  \frac{\sum\limits_{i=2}^N \! (x_i\!-\!x_{i-1})r_i}{N\!+\!\sum\limits_{i=2}^N x_ix_{i-1} } \!\right]  \!\right)\!,
 \end{aligned}
 \nonumber
\end{equation}
\begin{equation}
\begin{aligned}
\!\!\!\hat{h}_{1} \!=  \!\sqrt{\!\frac{\sigma_\eta^2}{2}}\! \!\left(\! \! {\rm erf}^{-1}  \!\!\left[ \! \frac{\sum\limits_{i=2}^N \! (x_i\!+\!x_{i-1})r_i}{N\!+\!\sum\limits_{i=2}^N x_ix_{i-1} } \!\right] \! \!+  \! {\rm erf}^{-1}  \! \!\left[ \!  \frac{\sum\limits_{i=2}^N \! (x_{i-1}\!-\!x_i)r_i}{N\!+\!\sum\limits_{i=2}^N x_ix_{i-1} } \!\right]  \!\right)\!.
 \end{aligned}
 \nonumber
\end{equation}
The solution consists in quite simple computations (apart of the final application of  ${\rm erf}^{-1}$)  since we only have to do with binary data ($r_i,x_i \in\{\pm 1\}$).
%%%%%%%%%%%%%%%%%%%%%%%%%%%%%%%%%%%%%%%%%%%%%%%%%%%%%%%%%%%%%%%%%%%%%%%%%%%%%%%%%%%%%%%%%%%%
\subsection{Non-Pilot Aided (Blind) Estimation}
\label{section:ExampleIII}
\begin{figure}[h]
\centerline{
\psfrag{CRB, MSE}[c][c]{\footnotesize  (MSE, CRB)/$h^2$}
\psfrag{SNR}[c][c]{ \footnotesize $h^2/\sigma_\eta^2$ (linear)} 
\epsfig{file=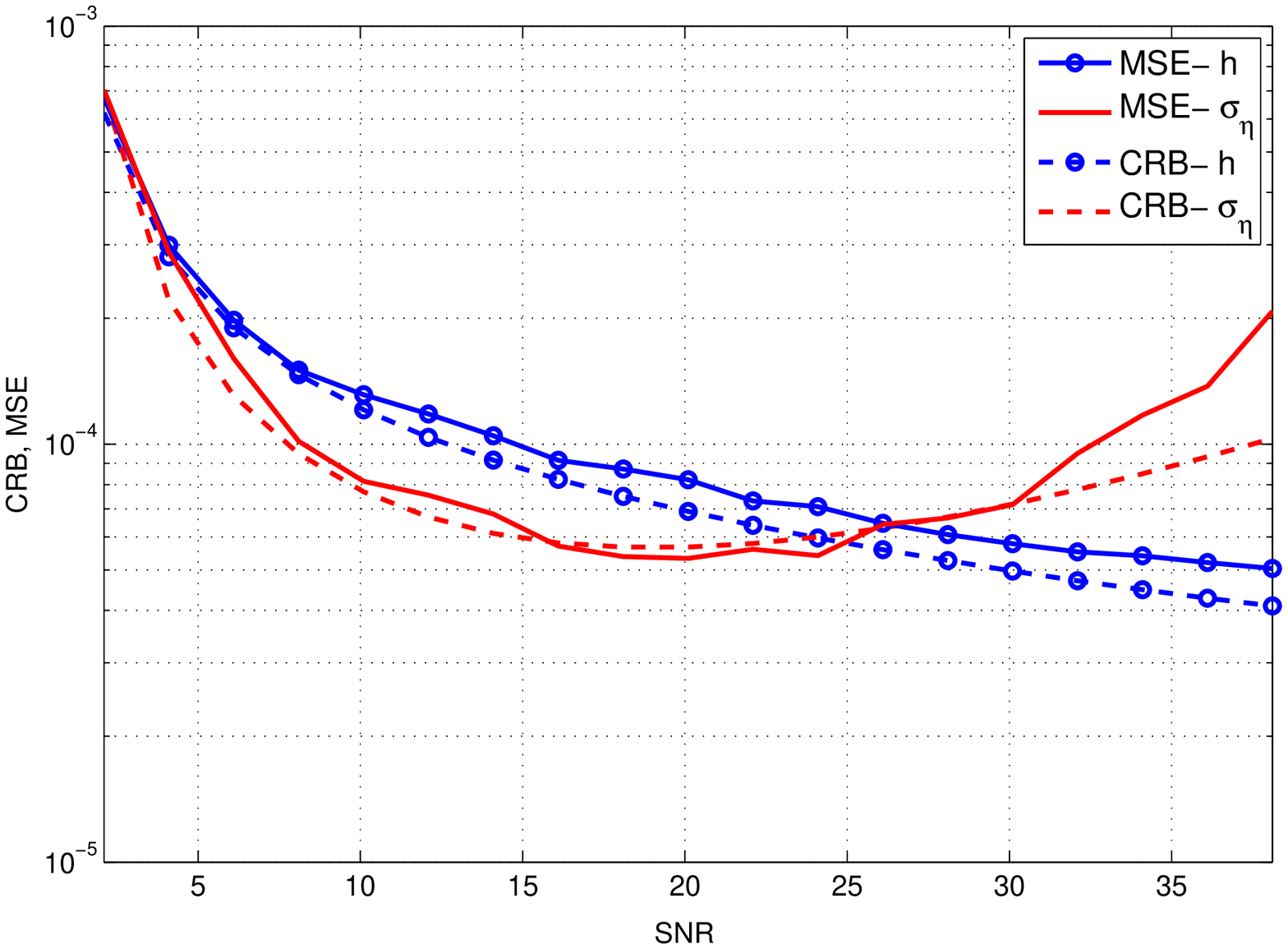, width =9cm}}
\caption{MSE and CRB of the blind estimates of $h$ and $\sigma_\eta$ vs. $h^2/\sigma_\eta^2$ for a SISO channel, $b=3$, $N=1000$.}
\label{fig:blind}
\end{figure}

%%%%%%%%%%%%%%%%%%%%%%%%%%%%%%%%%%%%%%%%%%%%%%%%%%%%%%%%%%%5%%%%%%%%%%%%%%%%%%%%%%%%%%%%%%%%%%%%%%%%%%
Suppose now that a unknown binary symbol sequence $x_i \in \{+1,.1\}$ is transmitted
over an additive white Gaussian noise (AWGN) channel with an unknown
real gain $h$. The analog channel output is
\begin{equation}
y_i=h\cdot x_i+ \eta,
\end{equation}
where the variance of the noise $\eta$, $\sigma_\eta^2$, is also unknown. Additionally, the receiver is unaware of the transmitted symbols $x_i$. Based on $N$ quantized observations $r_i = Q(y_i)$,
we wish to estimate the parameter vector $\B{\theta}=[h,\sigma_\eta]^{\rm T}$. We  note an inherent ambiguity in the
problem: the sign of the gain $h$ and the sign of $x_i$ cannot be determined
individually. We also note that at least 2 bits are needed in this case, because a single bit output does not contain any information about $h$. Since the ML problem is intractable in closed form, we resort to the EM approach. The EM-update for $h$ can be obtained from the general expressions in (\ref{estep}) and (\ref{mstep}) as
\begin{equation}
\begin{aligned}
&\hat{h}^{l+1}= \frac{1}{N}\sum_{i,x\in\{+1,-1\}}\\
&\frac{ \hat{h}^{l}[\Phi(\frac{r_i^{\rm up}\!-\!x \hat{h}^{l}  }{ \hat{\sigma}_\eta^l})\!-\!\Phi(\frac{r_i^{\rm lo}\!-\!x \hat{h}^{l}  }{ \hat{\sigma}_\eta^l})]\!-\!\frac{x\hat{\sigma}_\eta^l}{\sqrt{2\pi}} ({{\rm e}^{-\frac{(r_i^{\rm up}-x\hat{h}^{l})^2  }{ 2\hat{\sigma}_\eta^{l,2}}}\!\!-\!{\rm e}^{-\frac{(r_i^{\rm lo}\!-\!x \hat{h}^{l} )^2 }{ 2\hat{\sigma}_\eta^{l,2} }}} )}{\Phi(\frac{r_i^{\rm up}\!-\!\hat{h}^{l}  }{ \hat{\sigma}_\eta^l})-\Phi(\frac{r_i^{\rm lo}- \hat{h}^{l}  }{ \hat{\sigma}_\eta^l})+ \Phi(\frac{r_i^{\rm up}+ \hat{h}^{l}  }{ \hat{\sigma}_\eta^l})-\Phi(\frac{r_i^{\rm lo}+ \hat{h}^{l}  }{ \hat{\sigma}_\eta^l})}\\
&~~~~=\hat{h}^{l}-\frac{\hat{\sigma}_\eta^l}{\sqrt{2\pi}} \frac{1}{N} \sum_i\\
&~~~\frac{     {{\rm e}^{-\frac{(r_i^{\rm up}-\hat{h}^{l})^2  }{ 2\hat{\sigma}_\eta^{l,2}}}-{\rm e}^{-\frac{(r_i^{\rm lo}- \hat{h}^{l} )^2 }{ 2\hat{\sigma}_\eta^{l,2} }}} - {{\rm e}^{-\frac{(r_i^{\rm up}+\hat{h}^{l})^2  }{ 2\hat{\sigma}_\eta^{l,2}}}+{\rm e}^{-\frac{(r_i^{\rm lo}+ \hat{h}^{l} )^2 }{2 \hat{\sigma}_\eta^{l,2} }}}                                }{\Phi(\frac{r_i^{\rm up}-\hat{h}^{l}  }{ \hat{\sigma}_\eta^l})-\Phi(\frac{r_i^{\rm lo}- \hat{h}^{l}  }{ \hat{\sigma}_\eta^l})+ \Phi(\frac{r_i^{\rm up}+ \hat{h}^{l}  }{ \hat{\sigma}_\eta^l})-\Phi(\frac{r_i^{\rm lo}+ \hat{h}^{l}  }{ \hat{\sigma}_\eta^l})},
\end{aligned}
\end{equation}
while the update for the noise variance follows from the expectation
\begin{equation}
\begin{aligned}
&\hat{\sigma}^{l+1,2}_\eta=\frac{1}{N} \sum_i {\rm E}_{\B{x},\eta_i|r_i,h^l,\hat{\sigma}^{l,2}}[\eta_i^2]=\hat{\sigma}^{l,2}_\eta-\frac{\sqrt{2}\hat{\sigma}_\eta^l}{N\sqrt{\pi}}\sum_{i}\Bigg[\\
&\frac{     {(r_i^{\rm up}-\hat{h}^{l}){\rm e}^{-\frac{(r_i^{\rm up}-\hat{h}^{l})^2  }{ 2\hat{\sigma}_\eta^{l,2}}}-(r_i^{\rm lo}- \hat{h}^{l} ){\rm e}^{-\frac{(r_i^{\rm lo}- \hat{h}^{l} )^2 }{ 2\hat{\sigma}_\eta^{l,2} }}}                              }{\Phi(\frac{r_i^{\rm up}-\hat{h}^{l}  }{ \hat{\sigma}_\eta^l})-\Phi(\frac{r_i^{\rm lo}- \hat{h}^{l}  }{ \hat{\sigma}_\eta^l})+ \Phi(\frac{r_i^{\rm up}+ \hat{h}^{l}  }{ \hat{\sigma}_\eta^l})-\Phi(\frac{r_i^{\rm lo}+ \hat{h}^{l}  }{ \hat{\sigma}_\eta^l})}+\\
&\frac{     (r_i^{\rm up}+\hat{h}^{l}){\rm e}^{-\frac{(r_i^{\rm up}+\hat{h}^{l})^2  }{ 2\hat{\sigma}_\eta^{l,2}}}-(r_i^{\rm lo}+ \hat{h}^{l} ){\rm e}^{-\frac{(r_i^{\rm lo}+ \hat{h}^{l} )^2 }{ 2\hat{\sigma}_\eta^{l,2} }}                               }{\Phi(\frac{r_i^{\rm up}-\hat{h}^{l}  }{ \hat{\sigma}_\eta^l})-\Phi(\frac{r_i^{\rm lo}- \hat{h}^{l}  }{ \hat{\sigma}_\eta^l})+ \Phi(\frac{r_i^{\rm up}+ \hat{h}^{l}  }{ \hat{\sigma}_\eta^l})-\Phi(\frac{r_i^{\rm lo}+ \hat{h}^{l}  }{ \hat{\sigma}_\eta^l})} \Bigg],
\end{aligned}
\end{equation}
where we used the conditional distribution
\begin{equation}
\begin{aligned}
\!  p(\eta_i|r_i,\hat{\sigma}_\eta^l,\hat{h}^l)\!=\! \frac{     \mathbb{I}_{D(r_i)}\!(\eta_i\!+\!\hat{h}^{l})\frac{{\rm e}^{-\frac{\eta_i^2  }{ 2\hat{\sigma}_\eta^{l,2}}}}{\sqrt{\!2\pi\hat{\sigma}_\eta^{l,2}}}\!+\!\mathbb{I}_{D(r_i)}\!(\!\eta_i\!-\! \hat{h}^{l} )\frac{{\rm e}^{-\frac{\eta_i^2 }{ 2\hat{\sigma}_\eta^{l,2} }} }{\sqrt{\!2\pi\hat{\sigma}_\eta^{l,2}} }                           }{\Phi(\! \frac{r_i^{\rm up}-\hat{h}^{l}  }{ \hat{\sigma}_\eta^l}\!)\!-\!\Phi(\!\frac{r_i^{\rm lo}- \hat{h}^{l}  }{ \hat{\sigma}_\eta^l}\!)\!+\! \Phi(\!\frac{r_i^{\rm up}+ \hat{h}^{l}  }{\hat{\sigma}_\eta^l}\!)\!-\!\Phi(\!\frac{r_i^{\rm lo}+ \hat{h}^{l}  }{ \hat{\sigma}_\eta^l}\!)}\!.
\end{aligned}
\end{equation}
The Cram\'er-Rao Bound that can be easily obtained from the likelihood function
\begin{equation}
\begin{aligned}
\mathcal{L}(h,\sigma_\eta)\!&= \\
&\!\!\!\!\!\ln \! \left(\!\Phi(\! \frac{r_i^{\rm up}-{h}  }{ \sigma_\eta}\!)\!-\!\Phi(\!\frac{r_i^{\rm lo}- {h}  }{ \sigma_\eta}\!)\!+\! \Phi(\!\frac{r_i^{\rm up}+ {h}  }{ \sigma_\eta}\!)\!-\!\Phi(\!\frac{r_i^{\rm lo}+ {h}  }{ \sigma_\eta}\!) \!\right)\!\!,
\end{aligned}
\end{equation}
as well as the MSE of the estimates $\hat{h}$ and $\sigma_\eta$ found by Monte Carlo simulation, both normalized by $h^2$, is depicted in Fig.~\ref{fig:blind} as function of $h^2/\sigma_\eta^2$,  where the pilot length is $N=100$ and the quantizer resolution is $b=3$. Clearly the MSE of the estimate $\hat{\sigma}_\eta$ also exhibits the non-monotonic behavior mentioned before with respect to the SNR.

%\begin{equation}
%\begin{aligned}
% {(r_i^{\rm up}-\hat{h}^{l}){\rm e}^{-\frac{(r_i^{\rm up}-\hat{h}^{l})^2  }{ 2\sigma_\eta^{l,2}}}-(r_i^{\rm lo}- \hat{h}^{l} ){\rm %e}^{-\frac{(r_i^{\rm lo}- \hat{h}^{l} )^2 }{ 2\sigma_\eta^{l,2} }}}+(r_i^{\rm up}+\hat{h}^{l}){\rm e}^{-\frac{(r_i^{\rm up}+\hat{h}^{l})^2  }{ %2\sigma_\eta^{l,2}}}-(r_i^{\rm lo}+ \hat{h}^{l} ){\rm e}^{-\frac{(r_i^{\rm lo}+ \hat{h}^{l} )^2 }{ 2\sigma_\eta^{l,2} }} 
%\end{aligned}
%\end{equation}
%%%%%%%%%%%%%%%%%%%%%%%%%%%%%%%%%%%%%%%%%%%%%%%%%%%%%%%%%%%%%%%%%%%%%%%%%%%%%%%%%%%%%%%%%%%%%%%%%
%%%%%%%%%%%%%%%%%%%%%%%%%%%%%%%%%%%%%%%%%%%%%%%%%%%%%%%%%%%%%%%%%%%%%%%%%%%%%%%%%%%%%%%%%%%%%%%
%%%%%%%%%%%%%%%%%%%%%%%%%%%%%%%%%%%%%%%%%%%%%%%%%%%%%%%%%%%%%%%%%%%%%%%%%%%%%%%%%%%%%%%%
\section{Example 2: Pilot-based  MIMO channel Estimation}
\label{section:MIMO}
%%%%%%%%%%%%%%%%%%%%%%%%%%%%%%%%%%%%%%%%%%%%%%%%%%%%%%%%%%%%%%%%%%%%%%%%%%%%%%%%%%%%%%%%%%%%%%%
Now, we consider the MIMO case. We begin first with the problem of estimating a $2\times 2$ MIMO channel from single bit outputs, since it can be also solved in a closed form, as shown later on.
\subsection{Single bit estimation of a $2\times 2$ MIMO Channel}
\begin{figure}[h]
\centerline{
\psfrag{-log(ber)}[c][c]{\small{$-\mathrm{log}_\mathrm{e}$(BER)}}
\psfrag{Estimation Error}[c][c]{\footnotesize  MSE}
\epsfig{file=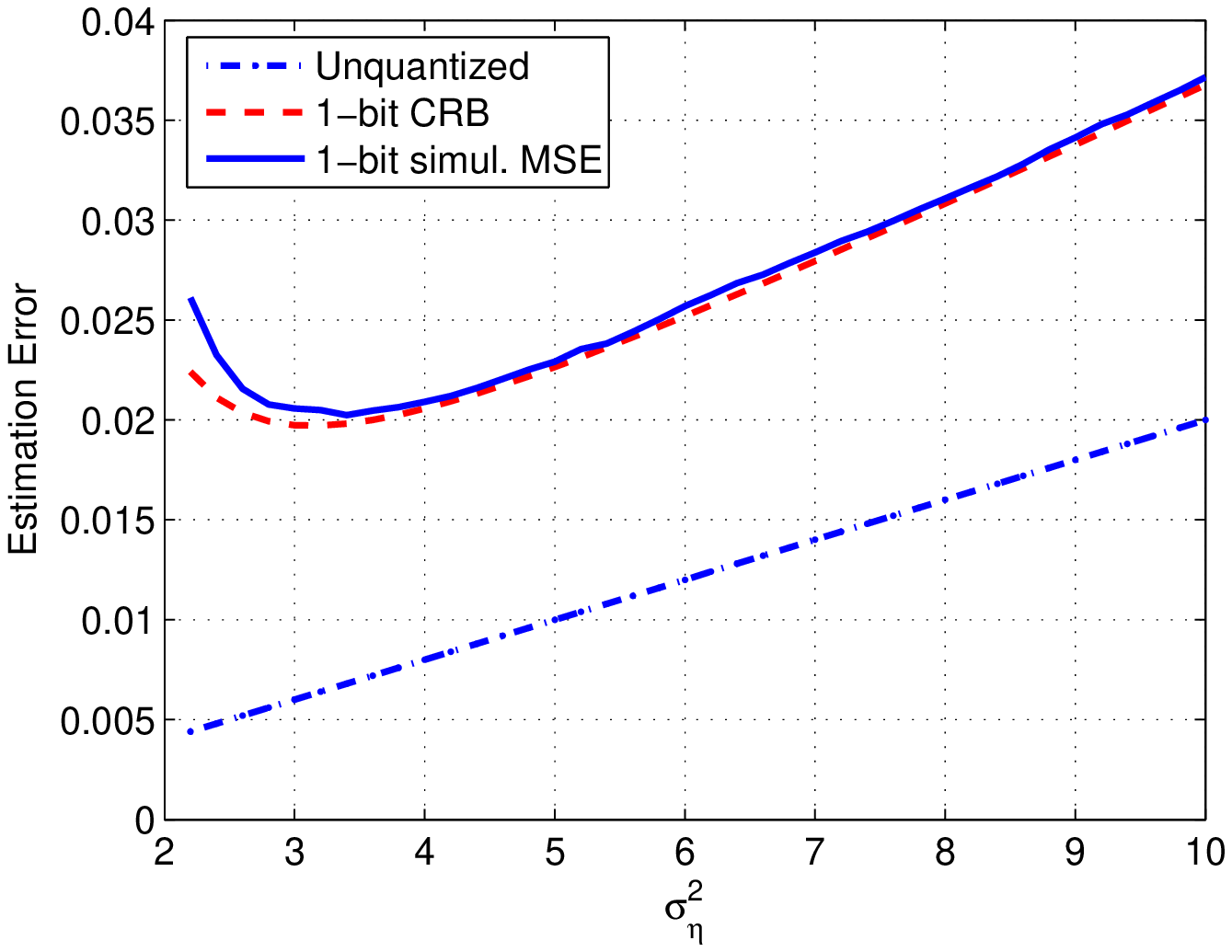, width =9cm}}
\caption{Estimation error vs. $\sigma_\eta^2$ for a 2$\times$2 real valued channel, $b=1$, $N=200$.}
\label{CRLB_fig}
\end{figure}

%%%%%%%%%%%%%%%%%%%%%%%%%%%%%%%%%%%%%%%%%%%%%%%%%%%%%%%%%%%5%%%%%%%%%%%%%%%%%%%%%%%%%%%%%%%%%%%%%%%%%%
As example, let us consider a pilot-based estimation of a real valued $2\times2$ channel matrix assuming a single-bit quantizer
\begin{equation}
\begin{aligned}
 &\B{r}_i= {\rm sign}( h_{i1}\B{x}_1+ h_{i2}\B{x}_2 + \B{\eta}_i), 
 \end{aligned}
\end{equation}
where $\B{x}_1$, $\B{x}_2\in \{-1,1\}^N$ are the pilot vectors transmitted at each Tx antenna, while $\B{r}_1$, $\B{r}_2\in \{-1,1\}^N$ are the received vectors at each Rx antenna. The maximum likelihood (ML) channel estimate $\hat{\B{\theta}}=[\hat{h}_{11},\hat{h}_{12},\hat{h}_{1},\hat{h}_{11}]^{\rm T}$ can be found by
solving (\ref{KKT}) in closed form, similarly to the 2-tap SISO channel (see Subsection~\ref{section:twotap}), as
\begin{equation}
\begin{aligned}
\hat{h}_{ij} \!=  \!\sqrt{\frac{\sigma_\eta^2}{2}} \!\left( \! {\rm erf}^{-1}  \!\!\left[ \frac{(\B{x}_1+\B{x}_2)^{\rm T} \B{r}_i}{N+\B{x}_1^{\rm T}\B{x}_2} \!\right]  \!+  \! {\rm erf}^{-1}  \! \!\left[ \! \frac{(\B{x}_j-\B{x}_{\bar j})^{\rm T} \B{r}_i}{N-\B{x}_1^{\rm T}\B{x}_2} \!\right]  \!\right),
 \end{aligned}
\end{equation}
with $i,j\in \{1,2\}$, ${\bar j}=3-j$ and $ {\rm erf}^{-1}$ the inverse function of the error function. We can see that the HW implementation of the estimation task is still considerably simple, since only shift registers, counters and a look-up table for $ {\rm erf}^{-1}$ would be necessary.
Fig.~\ref{CRLB_fig} shows the Monte Carlo simulation and the CRB of the estimation error $\sum_{ij}{\rm E}[(h_{ij}-\hat{h}_{ij})^2]$ for a given $2\times2$ channel as function of the noise variance $\sigma_\eta^2$. Thereby, the Fisher information matrix can be obtained from (\ref{CRLB}) as
\begin{equation}
\begin{aligned}
\B{J}=&   \frac{(N+\B{x}_1^{\rm T}\B{x}_2){\rm e}^{-\frac{(h_1+h_2)^2}{\sigma_\eta^2} }}{ 4\pi \sigma_\eta^2  \Phi(\frac{h_1+h_2}{\sigma_\eta} )\Phi(-\frac{h_1+h_2}{\sigma_\eta} )}
\left[\begin{array}{ll}
1&1 \\
1&1
\end{array}\right] + \\
& \frac{(N-\B{x}_1^{\rm T}\B{x}_2){\rm e}^{-\frac{(h_1-h_2)^2}{\sigma_\eta^2} }}{ 4\pi \sigma_\eta^2  \Phi(\frac{h_1-h_2}{\sigma_\eta} )\Phi(-\frac{h_1-h_2}{\sigma_\eta} )}
\left[\begin{array}{rr}
1&-1 \\
-1&1
\end{array}\right].
\end{aligned}
\end{equation}
As example, we took the specific channel matrix
\begin{equation}
\begin{aligned}
\B{H}=\left[ 
\begin{array}{cc}
	2 & 1.5 \\
	0.5 & -1
\end{array}
\right].
 \end{aligned}
\end{equation}
Fig.~\ref{CRLB_fig} shows also the MSE when using the analog (unquantized) output, which is exactly
\begin{equation}
\left.{\rm MSE}\right|_{b\rightarrow\infty}= \sigma_\eta^2 {\rm tr} \left(\left([\B{x}_1,\B{x}_2]^{\rm T}[\B{x}_1,\B{x}_2]\right)^{-1}\right).
\end{equation}
 Clearly, the estimation error under quantization does not increase monotonically with higher $\sigma_\eta^2$, as we know already from the SISO case.

%%%%%%%%%%%%%%%%%%%%%%%%%%%%%%%%%%%%%%%%%%%%%%
%%%%%%%%%%%%%%%%%%%%%%%%%%%%%%%%%%%%%%%%%%%%%%%%%%%%%%%%%%%%%%%%%%%%%%%%%%%%%%%%%%%%%%%%
\subsection{Pilot-based  MIMO channel estimation of arbitrary size}
\label{section:ExampleII}
%%%%%%%%%%%%%%%%%%%%%%%%%%%%%%%%%%%%%%%%%%%%%%%%%%%%%%%%%%%5%%%%%%%%%%%%%%%%%%%%%%%%%%%%%%%%%%%%%%%%%%
Let us consider now a more general setting of a quantized linear MIMO system
\begin{equation}
\B{y}={\rm vec} [\B{H}\B{X}']+\B{\eta},
\end{equation}
and 
\begin{equation}
\B{r}=Q(\B{y}),
\end{equation}
with a channel matrix $\B{H}\in \mathbb{R}^{L \times M}$ and $\B{X}' \in \mathbb{R}^{M\times N}$ contains $N$ pilot-vectors of dimension $M$. Hereby, we stack the unquantized, quantized and the noise signals into the vectors $\B{y}$, $\B{r}$ and $\B{\eta}$, respectively.
Our unknown parameter vector is therefore  $\B{\theta}=\B{h}={\rm vec}[\B{H}]$ and we have the system function
\begin{equation}
 \B{f}(\B{h},\B{X})= \B{X}\B{h},
\end{equation}
 where the new matrix $\B{X} \in \mathbb{R}^{{M\cdot N}\times {M\cdot L}}$ contains again the pilot-vectors in a proper way. Furthermore we assume, contrary to the previous cases, that  a priori  distribution $p(\B{h})$ is given according to $\B{h} \sim \mathcal{N} (\B{0}, \B{R}_h)$. With this definition the EM-iteration (\ref{estep}) and (\ref{mstep}) reads in this case as \vspace{0.5cm} \\
\underline{E-step:} Compute for $i=1,\ldots,N$
\begin{equation}
\begin{aligned}
b^{l}_i &= -\frac{\sigma_\eta}{\sqrt{2\pi}}\cdot \frac{{\rm e}^{-\frac{(r_{i}^{\rm up}- [\B{X}\hat{\B{h}}^{l}]_i )^2  }{ 2\sigma_\eta^2}}-{\rm e}^{-\frac{(r_{i}^{\rm lo}-[\B{X}\hat{\B{h}}^{l}]_i)^2 }{2 \sigma_\eta^2 }}}{\Phi(\frac{r_{i}^{\rm up}-[\B{X}\hat{\B{h}}^{l}]_i  }{ \sigma_\eta})-\Phi(\frac{r_{i}^{\rm lo}-[\B{X}\hat{\B{h}}^{l}]_i  }{ \sigma_\eta})} 
\end{aligned}
\end{equation}
\underline{M-step:} 
\begin{equation}
\begin{aligned}
 \hat{\B{h}}^{l+1}=(\B{X}^{\rm T}\B{X}+ \sigma_\eta^2 \B{R}^{-1}_h)^{-1}\cdot \B{X}^{\rm T} (\B{X} \hat{\B{h}}^{l} +\B{b}^{l} ).
\end{aligned}
\end{equation}
Let us at this point validate the convergence of the EM-algorithm to a unique optimum solution. For this, we write the log-likelihood function explicitly
\begin{align}
\mathcal{L}(\B{\theta})=\sum_i \!\ln \!\left( \!  \Phi(\frac{r_i^{\rm up}\!-\!\B{x}_i^{\rm T} \B{h}}{\sigma_\eta} )\!-\!\Phi(\frac{r_i^{\rm lo}-\B{x}_i^{\rm T} \B{h}}{\sigma_\eta} ) \! \right) \!- \!\B{h}^{\rm T} \B{R}^{-1} \B{h}.
\end{align}
This log-likelihood function is a smooth convex function with respect to $\B{\theta}$. This follows from the log-concavity of 
\begin{equation}
\Phi(b-z )-\Phi(a-z),
\end{equation}
 $b>a$, with respect to $z$, since it is obtained from the convolution of the Gaussian density and a normalized boxcar function localized  between $a$ an $b$, which are both log-concave \cite{boyd_convex}. 
Therefore, the stationary point of the EM-iteration fulfilling the condition (\ref{KKT}) is the unique optimal  solution.
\begin{figure}[h]
\centerline{
\psfrag{SNR}[c][c]{\small{$-10\mathrm{log}_{10}(\sigma_\eta^2)$}}
\psfrag{b} [c][c]{\small{$b$}} 
\epsfig{file=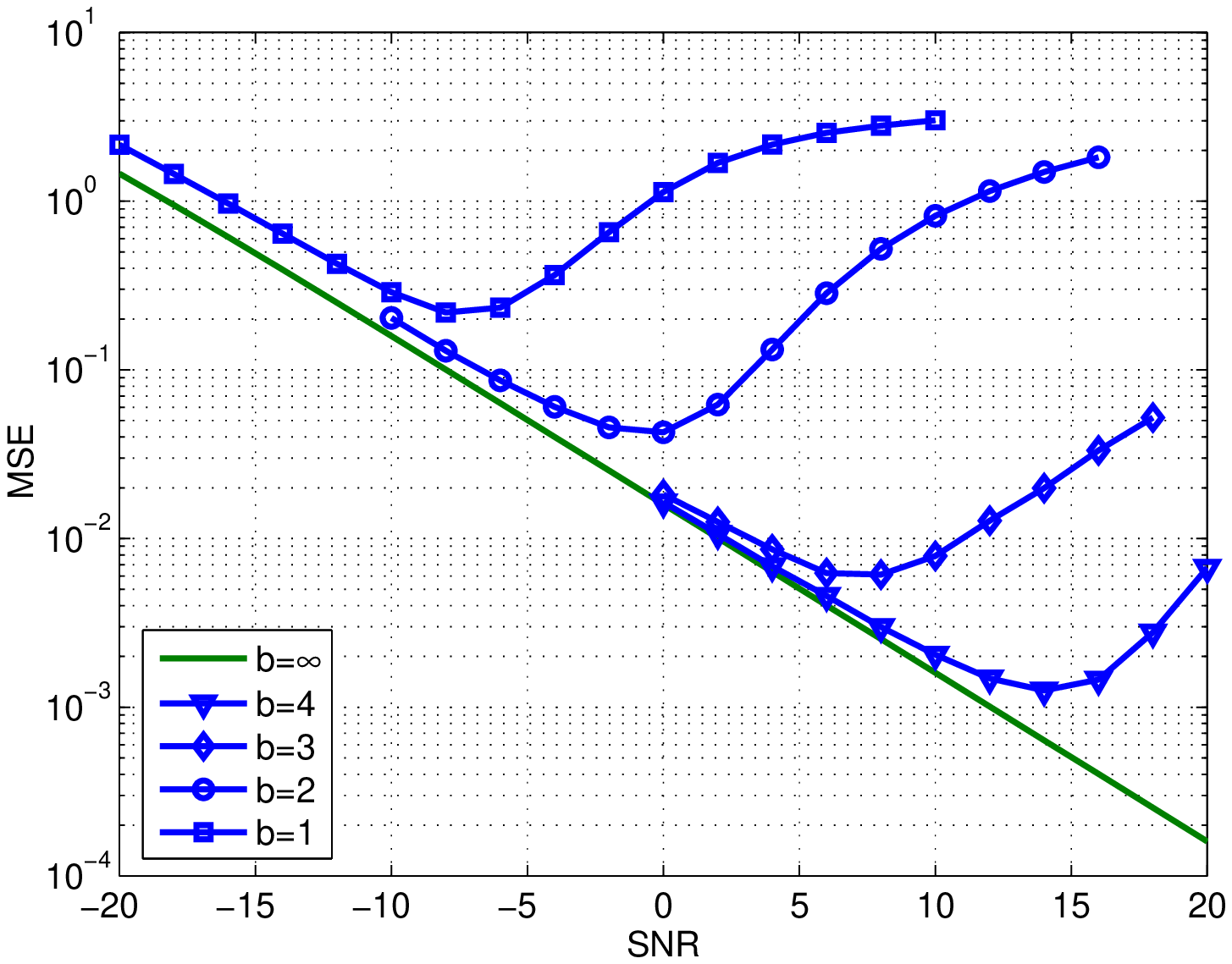, width =8.5cm}}
\caption{Estimation error vs. $\sigma_\eta^2$ for a real valued 4$\times$4 MIMO channel, $b=1$, $N=1000$, $\B{R}_h={\bf I}_{16}$, $x_{i,j}\in\{-1,+1\}$.}
\label{4x4MIMO_est}
\end{figure}

Fig.~\ref{4x4MIMO_est} illustrate the average MSE defined by
\begin{align}
{\rm MSE}={\rm E}\left[\left\| \B{h}-\hat{\B{h}}\right\|_2^2\right],
\end{align}
under different bit resolution for a 4$\times$4 MIMO channel with i.i.d. unit variance entries.
Hereby, we chose an orthogonal pilot sequence, i.e. $\B{X}^{\rm T}\B{X}=\B{R}_h={\bf I}_{16}$ with $x_{i,j}\in\{-1,+1\}$. The estimation error in the unquantized case, which is given by
\begin{align}
{\rm MSE}_{b \rightarrow \infty}= \sigma_\eta^2 {\rm tr}\left( (\B{X}^{\rm T}\B{X}+ \sigma_\eta^2 \B{R}^{-1}_h)^{-1}\right)
\end{align}
is also shown for comparison.  Obviously, at medium and low SNR, the coarse quantized does not affect the estimation performance considerably. 
%%%%%%%%%%%%%%%%%%%%%%%%%%%%%%%%%%%%%%%%%%%%%%%%%%%%%%%%%%%%%%%%%%%%%%%%%%%%%%%%%%%%%%%%%%%%
%%%%%%%%%%%%%%%%%%%%%%%%%%%%%%%%%%%%%%%%%%%%%%%%%%%%%%%%%%%%%%%%%%%%%%%%%%%%%%%%%%%%%%%%%%%%%%%%%
%%%%%%%%%%%%%%%%%%%%%%%%%%%%%%%%%%%%%%%%%%%%%%%%%%%%%%%%%%%%%%%%%%%%%%%%%%%%%%%%%%%%%%%%%%%%
%%%%%%%%%%%%%%%%%%%%%%%%%%%%%%%%%%%%%%%%%%%%%%%%%%%%%%%%%%%%%%%%%%%%%%%%%%%%%%%%%%%%%%%%%%%%
\section{Example 3: Quantization of GNSS Signals}
\label{section:GNSS}
The quality of the data provided by a GNSS receiver depends largely on the synchronization error with the signal transmitted by the 
GNSS satellite (navigation signal), that is, on the accuracy in the propagation time-delay estimation of the direct signal 
(line-of-sight signal, LOSS). In the following we will study the effect of quantization in terms of simulations and the CRB as derived in Section~\ref{SEC_CRB}. We assumed an optimal uniform quantizer as given in Table \ref{uniform}. We will first assess the accuracy of a standard one-antenna GNSS receiver in case no multipath is present. Secondly, we will assess the behavior of array synchronization signal processing in a multipath scenario applying the innovative derivation of the EM algorithm as shown in Section \ref{subsec:EM}. This assessment is based on the work presented in \cite{AnNoSeSw2009}. In the following we assume a GPS C/A code signal with a chip duration $T_c=977.52$ ns, a code length of $1$ ms and a bandwidth of $B=1.023$ MHz. The received signal is sampled with the sampling frequency $f_s=2 B$. We only use one code period as an observation time where the channel is assumed constant during this observation time.\par 
The synchronization of a navigation signal is usually performed by a Delay Lock Loop (DLL), which in case no multipath signals are present, efficiently implements a maximum likelihood estimator (MLE) for the time-delay of the LOSS $\tau_1$. 
%The RMSE of $\tau_1$ depending on the number of bits being used for quantization is depicted in Fig.~\ref{SISO_GNSS_del}.    
\begin{figure}[h!]
\centerline{
\psfrag{CRLB [Meter]}[B][c]{\small $\sqrt{\rm {CRB} (\hat{\tau}_1)}\cdot c_0$ [meter]}
\psfrag{SNR} [c][c]{\small SNR [dB]} 
\psfrag{b=1} [c][c]{\tiny $b=1$} 
\psfrag{b=2} [c][c]{\tiny $b=2$}
\psfrag{b=3} [c][c]{\tiny $b=3$}
\psfrag{b=4} [c][c]{\tiny $b=4$}
\epsfig{file=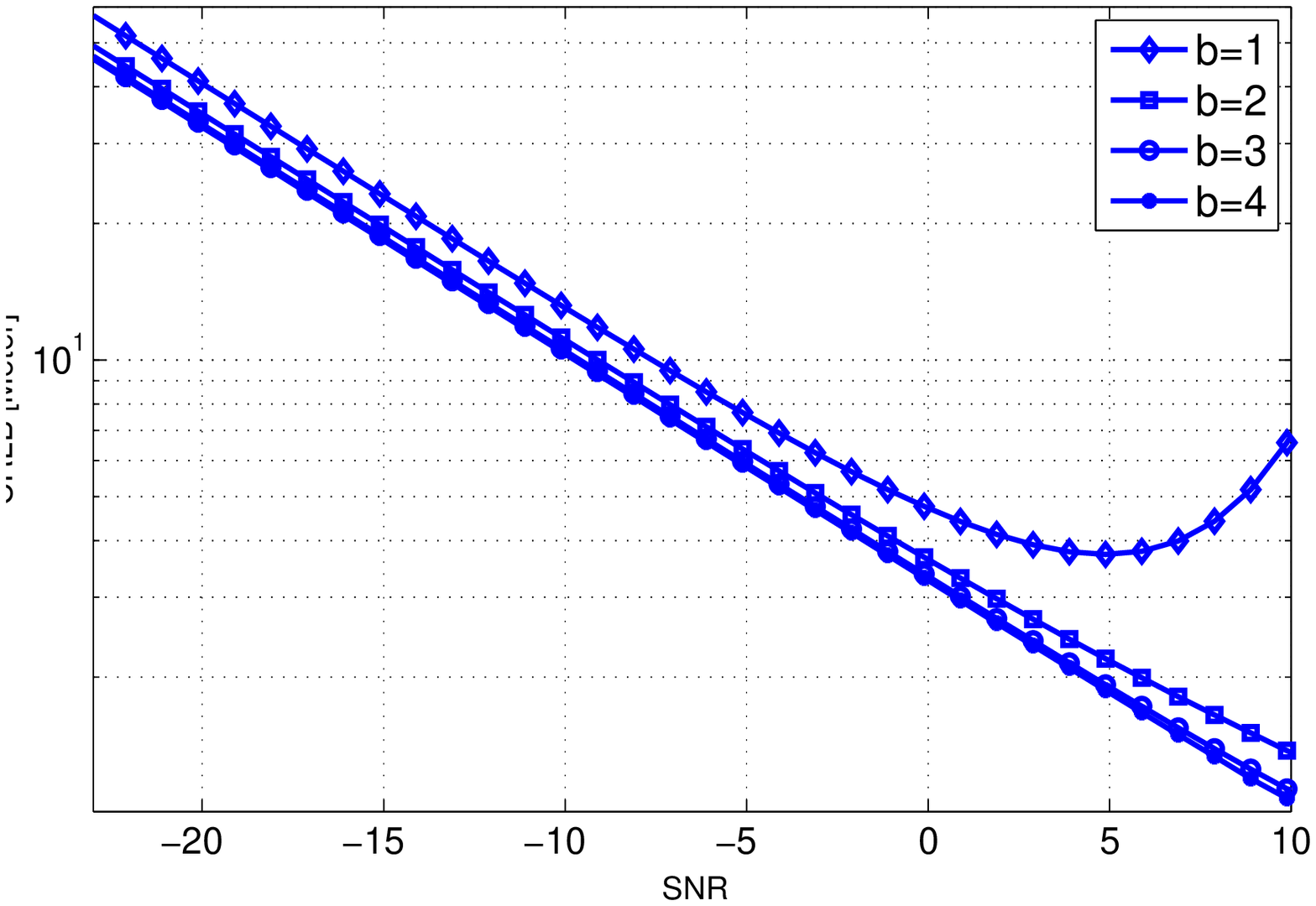, width =8.5cm}}
\caption{RMSE of $\hat{\tau}\cdot c_0$ vs.  bit resolution and SNR  with one antenna $f_s=2.046$MHz. One code period is used for estimation.}
\label{SISO_GNSS_del}
\end{figure}
In Fig.~\ref{SISO_GNSS_del}, the lower bound of the RMSE of $\tau_1$, for different the number of bits, is given in terms of the $\sqrt{\rm {CRB} (\hat{\tau}_1)}$ in meters where $c_0$ denotes the speed of light. A nominal SNR for a GPS C/A signal is approximately $-20$ dB. In Fig.~\ref{SISO_GNSS_del} one can observe that the $\sqrt{\rm {CRB} (\hat{\tau}_1)}$ does not significantly decrease further for more than 3 bits, thus a rather simple hardware implementation is sufficient for such a GNSS receiver.\par
\begin{figure}[h]
\centerline{
\psfrag{RMSE, CRLB}[B][c]{\small RMSE$\cdot c_0$, $\sqrt{\rm {CRB}}\cdot c_0$ [meter]}
\psfrag{bits} [c][c]{\small{$b$}}
\psfrag{CRB, LOSS}[c][c]{\tiny $\sqrt{\rm {CRB}(\hat{\tau}_1)}$}
\psfrag{CRB, MULT.}[c][c]{\tiny $\sqrt{\rm {CRB}(\hat{\tau}_2)}$}
\psfrag{RMSE, LOSS}[c][c]{\tiny $\rm{RMSE}(\hat{\tau}_1)$}
\psfrag{RMSE, MULT.}[c][c]{\tiny $\rm{RMSE}(\hat{\tau}_2)$}
\epsfig{file=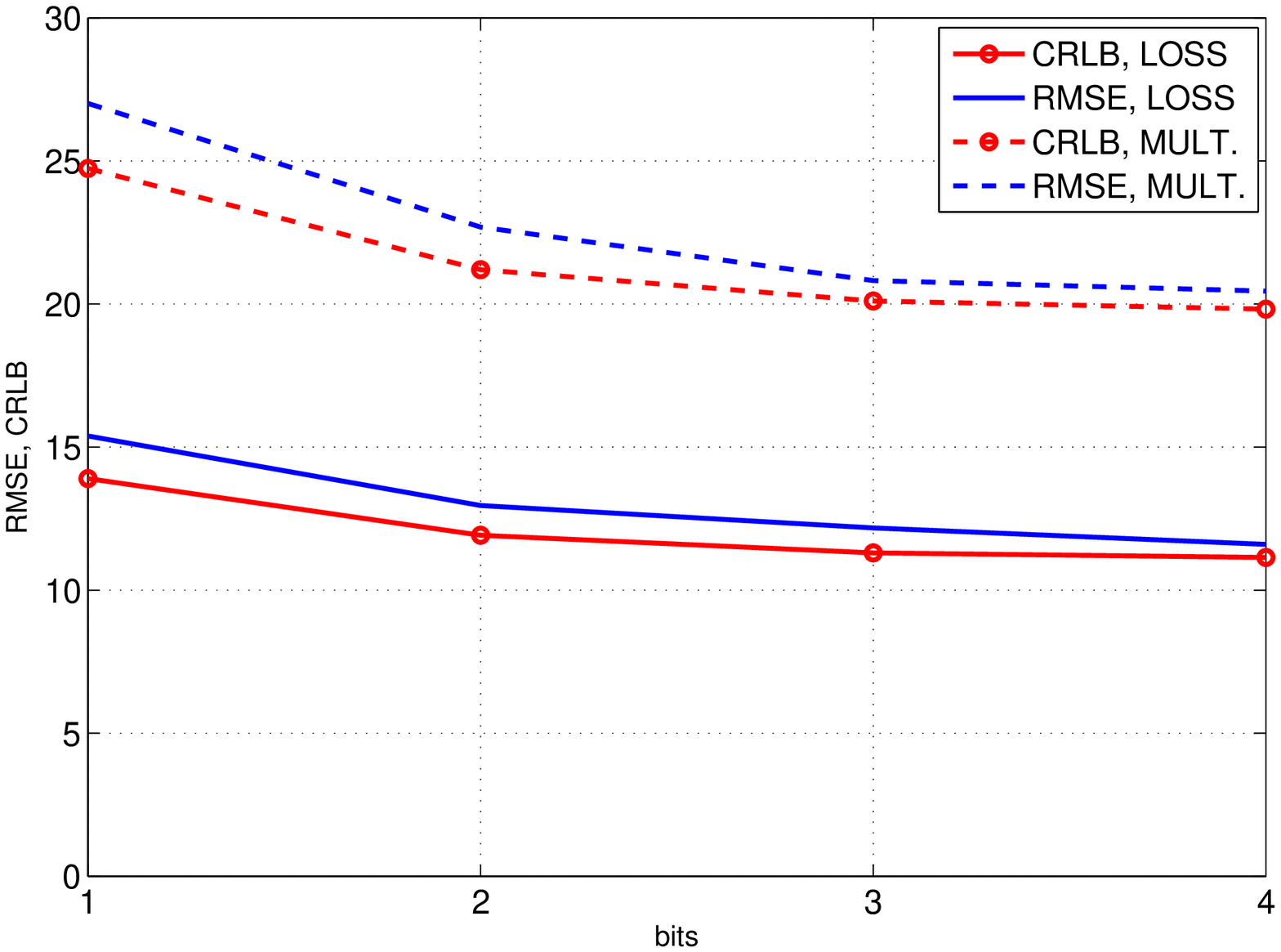, width =9cm}}
\caption{RMSE of $\hat{\tau}_{1}\cdot c_0$ and $\hat{\tau}_{2}\cdot c_0$ (in meter) vs.  bit resolution for $M=8$, $\phi_{1}=-30^\circ$, $\phi_{2}=62^\circ$,
 $\Delta \tau=0.3 T_c$, ${\rm SNR}=-22.8$dB, ${\rm SMR}=5$dB, $\Delta \nu=0$Hz. One code period is used for estimation.}
\label{8ant_GNSS_del}
\end{figure}

\begin{figure}[h]
\centerline{
\psfrag{RMSE, CRLB}[B][c]{\small RMSE, $\sqrt{\rm CRB}$ [$^{\circ}$]}
\psfrag{bits} [c][c]{\small{$b$}} 
\psfrag{CRLB, LOSS}[c][c]{\tiny $\sqrt{\rm {CRB}(\hat{\phi}_1)}$}
\psfrag{CRLB, MULT.}[c][c]{\tiny $\sqrt{\rm {CRB}(\hat{\phi}_2)}$}
\psfrag{RMSE, LOSS}[c][c]{\tiny $\rm{RMSE}(\hat{\phi}_1)$}
\psfrag{RMSE, MULT.}[c][c]{\tiny $\rm{RMSE}(\hat{\phi}_2)$}
\epsfig{file=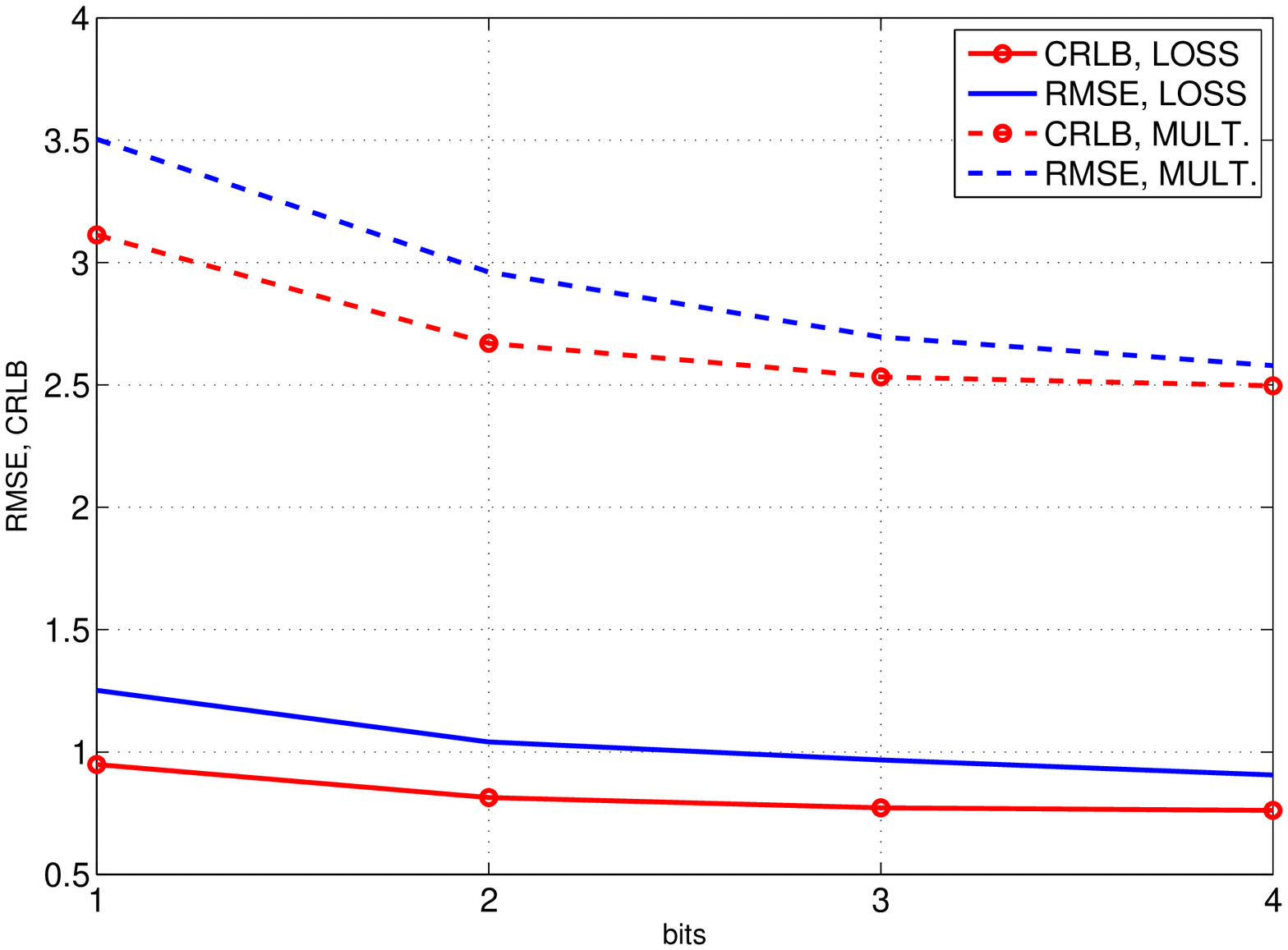, width =9cm}}
\caption{RMSE of $\hat{\phi}_{1}$ and  $\hat{\phi}_{2}$ (in degree) vs.  bit resolution for $M=8$, $\phi_{2}=-30^\circ$, $\phi_{2}=62^\circ$, $\Delta \tau=0.3 T_c$, ${\rm SNR}=-22.8$dB, ${\rm SMR}=5$dB, $\Delta \nu=0$Hz. One code period is used for estimation.}
\label{8ant_GNSS_azi}
\end{figure}

  
Now, we assess the EM algorithm as derived in Section \ref{subsec:EM} with $p_{\theta}(\B{\theta})$ being a uniform distribution, hence considering a ML estimator. We consider a two path scenario where the LOSS and one reflective multipath signal are received by an uniform linear antenna array (ULA) with $M=8$ isotropic sensor elements. We define
\begin{equation}
\B{\theta}=[\rm{Re}\{\B{\gamma}\}^{\rm T},\rm{Im}\{\B{\gamma}\}^{\rm T},\B{\tau}^{\rm T},\B{\nu}^{\rm T},\B{\phi}^{\rm T}]^{\rm T},
\end{equation}
with the vector of complex amplitudes $\B{\gamma}=[\gamma_{1},\gamma_{2}]^{\rm{T}}$, the vector of azimuth angles $\B{\phi}=[\phi_{1},\phi_{2}]^{\rm{T}}$, the vector of time-delays
$\B{\tau}=[\tau_{1},\tau_{2}]^{\rm{T}}$, and the vector of Doppler frequencies
$\B{\nu}=[\nu_{1},\nu_{2}]^{\rm{T}}$. The parameters with the subscript 1 refer to the LOSS and parameters with the subscript 2 refer to the reflection. The reflected multipath and the LOSS are considered to be in-phase, which means $\arg (\gamma_1) = \arg (\gamma_2) $, and the signal-to-multipath ratio (SMR) is 5dB. Signal-to-noise ratio (SNR) denotes the LOSS-to-noise ratio and we assume $\rm{SNR}=-22.8$dB. The DOAs for the LOSS and the multipath are $\phi_{1}=-30^{\circ}$ and $\phi_{2}=62^{\circ}$ respectively.  Further, we define the relative time-delay between the LOSS and the multipath as $\Delta\tau=|\tau_1-\tau_2|=0.3 T_c$ and relative Doppler $\Delta \nu=|\nu_1-\nu_2|=0$Hz. In Fig.~\ref{8ant_GNSS_del} the RMSE of $\hat{\tau}_{1}$ and $\hat{\tau}_{2}$ vs. the bit resolution is depicted.  
In Fig.~\ref{8ant_GNSS_azi} the RMSE of $\hat{\phi}_{1}$ and  $\hat{\phi}_{2}$ vs. the bit resolution is shown.
Based on the results presented in Fig.~\ref{8ant_GNSS_del} and Fig.~\ref{8ant_GNSS_azi} one can derive the important statement that 4 bits seem to be sufficient for high-resolution estimates with respect to the considered channel conditions.
%%%%%%%%%%%%%%%%%%%%%%%%%%%%%%%%%%%%%%%%%%%%%%%%%%%%%%%%%%%5%%%%%%%%%%%%%%%%%%%%%%%%%%%%%%%%%%%%%%%%%%
%%%%%%%%%%%%%%%%%%%%%%%%%%%%%%%%%%%%%%%%%%%%%%%%%%%%%%%%%%%%%%%%%%%%%%%%%%%%%%%%%%%%%%%%%%%%%%%%%
%%%%%%%%%%%%%%%%%%%%%%%%%%%%%%%%%%%%%%%%%%%%%%%%%%%%%%%%%%%%%%%%%%%%%%%%%%%%%%%%%%%%%%%%%%%%%%%%%
%%%%%%%%%%%%%%%%%%%%%%%%%%%%%%%%%%%%%%%%%%%%%%%%%%%%%%%%%%%%%%%%%%%%%%%%%%%%%%%%%%%%%%%%%%%%%%%%%
\section{Conclusion}
A general EM-based approach for  optimal parameter estimation based on quantized channel outputs has been presented. It has been applied in channel estimation and for GNSS.  Besides, the performance limit given by the Cram\'er-Rao Bound (CRB) has been discussed as well as  the effects of quantization and the optimal choice of the ADC characteristic. It turns out that the gap to the ideal (infinite precision) case  in terms of estimation performance is relatively small especially at low SNR. This holds independently of whether the quantizer is uniform or not. Additionally, we observed that the additive noise  might, at certain level, be favorable when operating on quantized data, since the MSE curves that we obtained were not monotonic with the SNR. This is an interesting phenomenon that could be investigated in future works.

\bibliographystyle{IEEEbib}
\setlength{\textheight}{16.5 cm}
\bibliography{IEEEabrv,references}
%%%%%%%%%%%%%%%%%
\end{document}